\documentclass[11pt,a4paper,pdftex]{article}
\pdfoutput=1
\usepackage{jcappub}

\newcommand{\be}{\begin{eqnarray}}
\newcommand{\ee}{\end{eqnarray}}
\newcommand{\rar}{\rightarrow}

\title{Measuring the Kerr spin parameter of regular black holes from their shadow}

\author{Zilong Li and Cosimo Bambi$^1$ 
\note{Corresponding author}}

\affiliation{Center for Field Theory and Particle Physics \& Department of Physics,\\
Fudan University,\\
220 Handan Road, 200433 Shanghai, China}

\emailAdd{zilongli@fudan.edu.cn}
\emailAdd{bambi@fudan.edu.cn}

\abstract{In a previous paper, one of us has showed that, at least in some cases,
the Kerr-nature of astrophysical black hole candidates is extremely difficult to test 
and current techniques, even in presence of excellent data not available today, 
cannot distinguish a Kerr black hole from a Bardeen one, despite the substantial 
difference of the two backgrounds. In this paper, we investigate if the detection of 
the ``shadow'' of nearby super-massive black hole candidates by near future 
mm/sub-mm very long baseline interferometry experiments can do the job. More 
specifically, we consider the measurement of the Kerr spin parameter of the Bardeen 
and Hayward regular black holes from their shadow, and we then compare the 
result with the estimate inferred from the K$\alpha$ iron line and from the frequency 
of the innermost stable circular orbit. For non-rotating black holes, the shadow 
approach provides different values, and therefore the Kerr black hole hypothesis can 
potentially be tested. For near extremal objects, all the approaches give quite 
similar results, and therefore it is not possible to constrain deviations from the
Kerr solution. The present work confirms that it is definitively challenging to test 
this kind of metrics, even with future facilities. However, the detection of a source 
that looks like a fast-rotating Kerr black hole can put meaningful constraints on 
the nature of the compact object.}


\begin{document}

\maketitle


\section{Introduction}

In 4-dimensional general relativity, uncharged black holes (BHs) are described
by the Kerr solution and are completely specified by two parameters, the mass
$M$ and the spin angular momentum $J$~\cite{bhth-p1,bhth-p2,bhth-p3}. The 
condition for the
existence of the event horizon is $|a_*| \le 1$, where $a_* = a/M = J/M^2$ is the 
spin parameter\footnote{Throughout the paper, we use units in which $G_{\rm N}
= c = 1$, unless stated otherwise.}. Astrophysical BHs, if they exist, are expected
to be well described by the Kerr metric: initial deviations from the Kerr geometry
should be quickly radiated away through the emission of gravitational
waves~\cite{price-p1,price-p2}, an initially non-vanishing electric charge would be 
shortly neutralized in their highly ionized environment~\cite{bdp}, while the presence 
of the accretion disk is completely negligible in most cases. Astrophysical BH 
candidates are dark compact objects in X-ray binary systems with a mass 
$M \approx 5 - 20$~$M_\odot$ and super-massive bodies 
in galactic nuclei with a mass $M \sim 10^5 - 
10^9$~$M_\odot$~\cite{nara}. They are thought to be the Kerr BHs of general 
relativity, but their actual nature is still to be verified. Stellar-mass BH candidates 
are simply too heavy to be neutron or quark stars for any plausible matter 
equation of state~\cite{bh1-p1,bh1-p2}. At least some of the super-massive BH candidates 
at the centers of galaxies are too massive, compact, and old to be clusters of 
non-luminous bodies~\cite{bh2}. The non-observation of electromagnetic radiation 
emitted by the possible surface of these objects may also be interpreted as an 
indication for the existence of an event horizon~\cite{horizon1-p1,horizon1-p2} 
(but see~\cite{horizon2-p1,horizon2-p2}). 
However, there is no evidence that the spacetime geometry around them is 
really described by the Kerr solution.

The possibility of testing the nature of astrophysical BH candidates with current
and near future observations has recently become a quite active research field
(for a review, see~\cite{review-p1,review-p2}). Today, there are two relatively robust techniques
to estimate the spin parameter of BH candidates under the assumption that the
geometry around them is described by the Kerr metric: the so-called continuum-fitting
method~\cite{cfm-p1,cfm-p2,cfm-p3} and the analysis of the K$\alpha$ iron 
line~\cite{iron-p1,iron-p2,iron-p3}. Both
the approaches can be used to probe the geometry of the spacetime around
BH candidates and measure the spin parameter and possible deviations from
the Kerr solution~\cite{torres-p1,torres-p2,cb-cfm-p1,cb-cfm-p2,cb-iron-p1,cb-iron-p2,cb-iron-p3,cb-iron-p4}. However, it turns out that there is a strong
correlation between the spin and possible deformations and that one can only
constrain a certain combination of these quantities. In other words,
the thermal spectrum of a thin accretion disk and the profile of the K$\alpha$
iron line of a Kerr BH with spin parameter $a_*$ can be extremely similar --
practically indistinguishable -- from the ones of non-Kerr compact objects with
different spin parameters. In Ref.~\cite{cb-ci}, one of us has showed that, at least
for some non-Kerr metrics, the combination of the continuum-fitting method 
and of the iron line analysis cannot fix this problem. Other approaches to test the 
nature of BH candidates are either not yet mature, like the case of quasi-periodic
oscillations~\cite{qpo-p1,qpo-p2}, or it is not clear when astrophysical data will be available,
like the case of gravitational waves or observations of a BH binary with a pulsar
companion~\cite{future-p1,future-p2,future-p3,future-p4,future-p5,future-p6,future-p7}. 
The estimate of the power of steady and transient jets
can potentially break the degeneracy between spin parameter and deviations
from the Kerr solution~\cite{jets-p1,jets-p2}, but at present we do not know the exact 
mechanism responsible for these phenomena. A rough estimate of possible
deviations from the Kerr geometry in the spacetime around super-massive BH
candidates can be obtained from considerations on their radiative efficiency
and on the possible mechanisms capable of spinning them up and
down~\cite{cb-bound-p1,cb-bound-p2,cb-bound-p3,cb-bound-p4}.

A quite promising technique to test the nature of super-massive BH candidates
with near future very long baseline interferometry (VLBI) facilities is through the
observation of the ``shadow'' of these objects~\cite{vlbi-p1,vlbi-p2}. The shadow is a dark
area over a bright background appearing in the image of an optically thin emitting
region around a BH~\cite{book,shadow-p1,shadow-p2}. 
While the intensity map of the image depends
on the details of the accretion process and of the emission mechanisms, the
boundary of the shadow is only determined by the metric of the spacetime, since
it corresponds to the apparent image of the photon capture sphere as seen by a
distant observer. The possibility of testing the nature of supermassive BH candidates
by observing the shape of their shadow has been already discussed in the literature,
starting from Ref.~\cite{sh-1-p1,sh-1-p2}. In general, very accurate observations are necessary,
because the effect of possible deviations from the Kerr solution are 
tiny~\cite{sh-2-p1,sh-2-p2,sh-2-p3,sh-2-p4,sh-2-p5,sh-2-p6,sh-2-p7}.

At first approximation, the shape of the shadow of a BH is a circle. The radius
of the circle corresponds to the apparent photon capture radius, which, for a given
metric, is set by the mass of the compact object and its distance from us. These two
quantities are usually known with a large uncertainty, and therefore the observation
of the size of the shadow can unlikely be used to test the nature of the BH candidate
(but see Ref.~\cite{sh-2-p7}). The shape of the shadow is instead the key-point.
The first order correction to the circle is due to the BH spin, as the photon
capture radius is different for co-rotating and counter-rotating particles. The
boundary of the shadow has thus a dent on one side: the deformation is more pronounced
for an observer on the equatorial plane (viewing angle $i = 90^\circ$) and decreases
as the observer moves towards the spin axis, to completely disappear when
$i = 0^\circ$ or $180^\circ$. Possible deviations from the Kerr solutions usually
introduces smaller corrections.

In the present paper, we consider the measurement of the Kerr spin parameter
of Kerr BHs and non-Kerr regular BHs; that is, we measure the spin parameter $a_*$ from
the shape of the shadow of a BH assuming it is of the Kerr kind. We use the
procedure proposed in Ref.~\cite{maeda}, which is based on the determination
of the distortion parameter $\delta_s = D_{cs}/R_s$, where $D_{cs}$ and
$R_s$ are, respectively, the dent and the radius of the shadow.
In the case of non-Kerr BHs, this
technique provides the correct value of $a_*$ for non-rotating objects, but
a quite different spin for near extremal states. We then compare these
measurements with the ones we could infer from the analysis of the K$\alpha$
iron line and the observation of a hot spot orbiting around the BH candidate.
The K$\alpha$ iron line approach is currently the only relatively robust
technique to probe the spacetime geometry around these objects. The
observations of hot spots orbiting around nearby super-massive BH candidates
with mm/sub-mm VLBI facilities will hopefully allow to determine the
frequency of test-particles at the innermost stable circular orbit (ISCO) radius.
For non-rotating and slow-rotating objects, the shadow approach provides
different results with respect to the other two techniques, so that a possible
combination of these methods may break the degeneracy between spin
parameter and possible deviations from the Kerr solution. All the approaches
seem instead to provide quite similar measurements for near extremal BHs,
which means that their combination cannot be used to test the spacetime
geometry around these objects. Our work confirms the difficulty
to observationally test the Kerr nature of astrophysical BH candidates.
Only very good observations of the shadow, which are capable of measuring
simultaneously the spin and possible deformations from the Kerr solution,
might be required to test the Kerr BH hypothesis. However, in the case
of objects that look like very fast-rotating Kerr BHs, interesting constraints on 
their nature seem to be possible.

The content of the paper is as follows. In Section~\ref{s-s}, we present our
approach to test the nature of astrophysical BH candidates and we introduce
the metrics that will be used in the rest of the paper. In Section~\ref{s-s2}, we 
briefly review the concept and the calculation of the BH's shadow. Section~\ref{s-m}
is devoted to the measurement of the Kerr spin parameter: we apply the procedure 
proposed in Ref.~\cite{maeda} to measure the spin of a Kerr BH to the Bardeen 
and Hayward BHs. Such a prescription provides the correct value of the spin 
parameter for non-rotating objects, but a wrong estimate for fast-rotating BHs. 
The results are then compared with the measurements we would obtain from 
the analysis of the K$\alpha$ iron line and the hot spot model in Section~\ref{s-m2}. 
Summary and conclusions are in Section~\ref{s-c}.

\section{Testing the Kerr-nature of black hole candidates \label{s-s}}

In Boyer-Lindquist coordinates, the non-vanishing metric coefficients of the
Kerr metric are
\be\label{eq-metric}
&&g_{tt} = - \left(1 - \frac{2 M r}{\Sigma}\right) \, , \quad
g_{t\phi} = - \frac{2 a M r \sin^2 \theta}{\Sigma} \, , \nonumber\\
&&g_{\phi\phi} = \left(r^2 + a^2 + \frac{2 a^2 M r \sin^2\theta}{\Sigma}\right)
\sin^2\theta \, , \quad
g_{rr} = \frac{\Sigma}{\Delta}\, ,
\quad g_{\theta\theta} = \Sigma \, ,
\ee
where
\be
\Sigma = r^2 + a^2 \cos^2\theta \, , \quad
\Delta = r^2 - 2 M r + a^2 \, .
\ee
$M$ is the BH mass and $a = J/M$ is its spin parameter.
If we want to test the Kerr nature of an astrophysical BH candidate, it is
convenient to consider a more general spacetime, in which the central object is
described by a mass $M$, spin parameter $a$, and one (or more)
``deformation paramater(s)''. The latter measure possible deviations from the
Kerr solution, which must be recovered when all the deformation parameters
vanish. The strategy is thus to calculate some observables in 
this more general background and then
fit the data of the source to find the allowed values of the
spin and of the deformation parameters. If the observations require vanishing
deformation parameters, the compact object is a Kerr BH. 
If they demand non-vanishing deformation parameters,
astrophysical BH candidates are not the Kerr BH of general relativity and new
physics is necessary. In general, however, the result is that observations
allows both the possibility of a Kerr BH with a certain spin parameter and
non-Kerr objects with different spin parameters.

\begin{figure}
\begin{center}
\hspace{-1.5cm}
\includegraphics[type=pdf,ext=.pdf,read=.pdf,width=9.5cm]{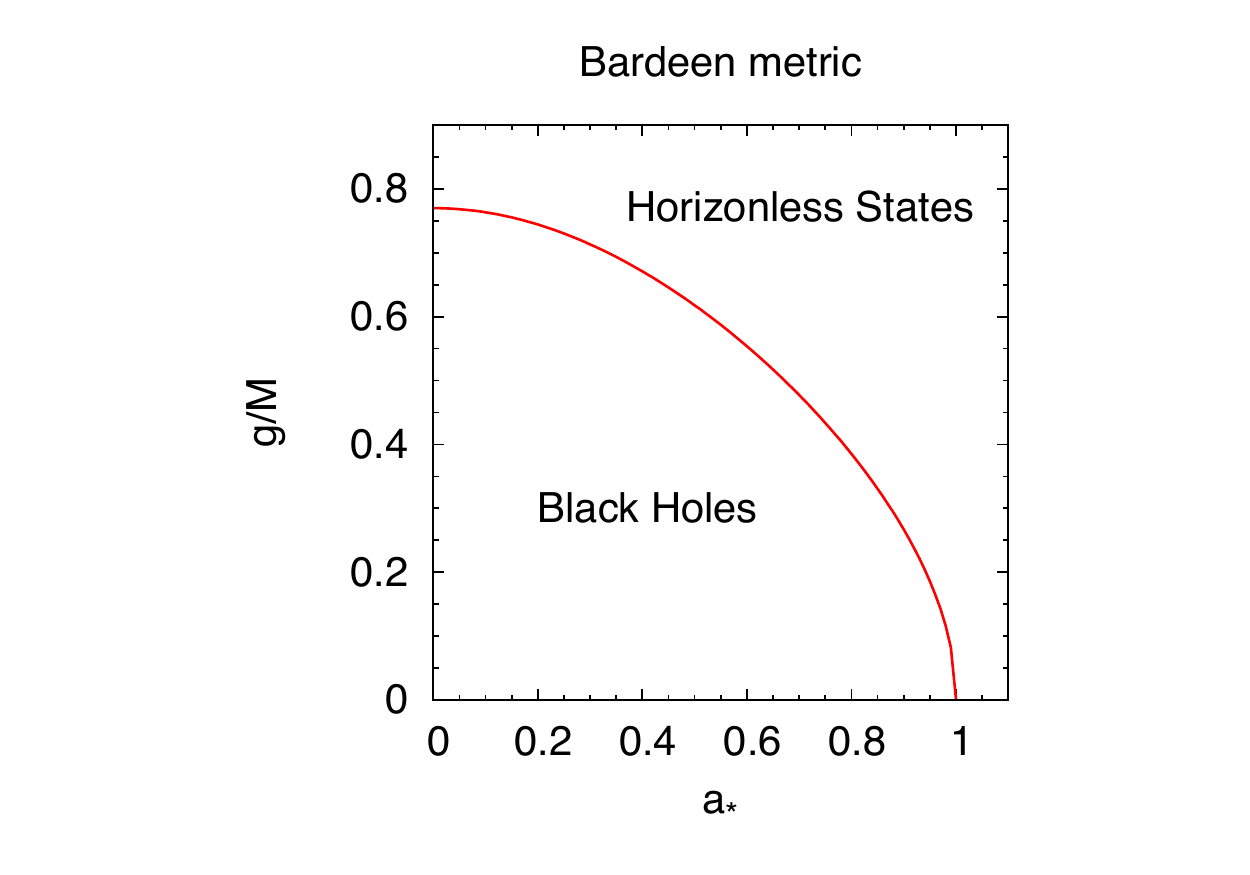}
\hspace{-3cm}
\includegraphics[type=pdf,ext=.pdf,read=.pdf,width=9.5cm]{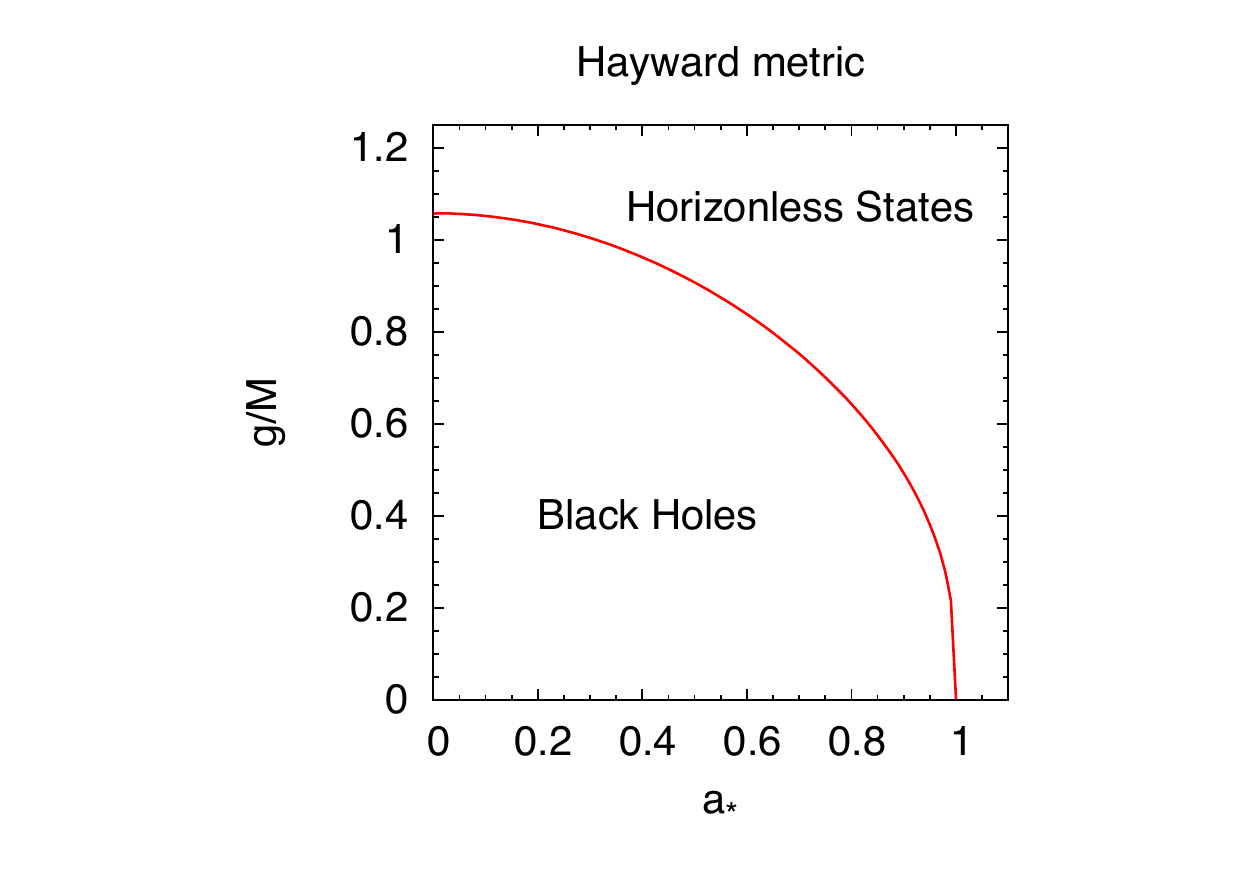}
\end{center}
\vspace{-0.5cm}
\caption{Rotating Bardeen (left panel) and Hayward (right panel) metrics. 
The red solid lines separate the regions
with BHs from the ones with horizonless objects.}
\label{fig00}
\end{figure}

As non-Kerr metrics, in the present work we will focus on the Bardeen and 
Hayward BHs~\cite{regularBHs-p1,regularBHs-p2}, which cannot be observationally tested by
current techniques (continuum-fitting method and iron line analysis), even 
in presence of excellent data not available today~\cite{cb-ci}. The rotating
solutions have the same form of the Kerr metric, with the mass $M$ replaced 
by $m$ as follows~\cite{regular-p1,regular-p2}:
\be
M &\rar& m_{\rm B} = M \left(\frac{r^2}{r^2 + g^2}\right)^{3/2} \, ,  \label{m-bardeen} \\
M &\rar& m_{\rm H} = M \frac{r^3}{r^3 + g^3} \, .  \label{m-hayward}
\ee
$g$ can be interpreted as the magnetic charge of a non-linear electromagnetic
field or just as a quantity introducing a deviation from the Kerr metric and solving
the central singularity. The position of the even horizon is given by the larger 
root of $\Delta = 0$ and therefore there is a bound on the maximum value of the 
spin parameter, above which there are no BHs. The maximum value of $a_*$ is 
1 for $g/M = 0$ (Kerr case), and decreases as $g/M$ increases. 
The regions of BHs and horizonless states on the plane $(a_*, g/M)$
are shown in Fig.~\ref{fig00}. In what follows, we will restrict the attention to the BH
region: even if they can be created~\cite{regular-p2}, 
the horizonless states are likely very 
unstable objects with a short lifetime due to the ergoregion instability.

\begin{figure}
\begin{center}
\includegraphics[type=pdf,ext=.pdf,read=.pdf,width=6cm]{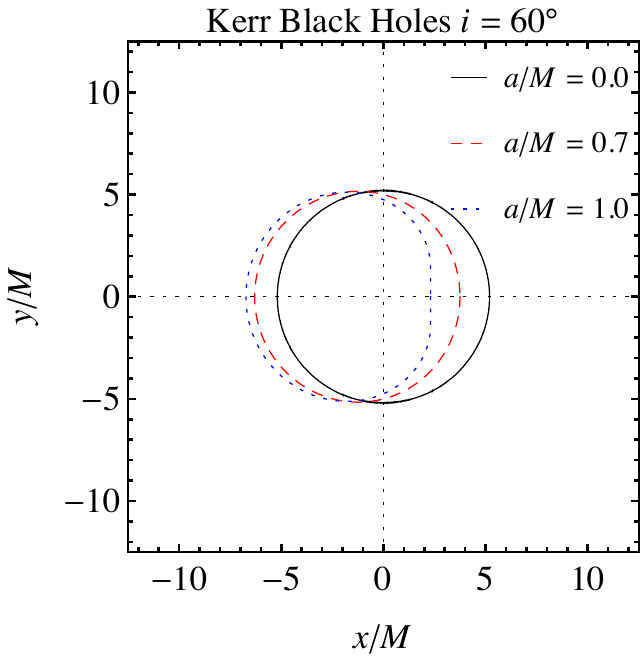}
\hspace{0.5cm}
\includegraphics[type=pdf,ext=.pdf,read=.pdf,width=6cm]{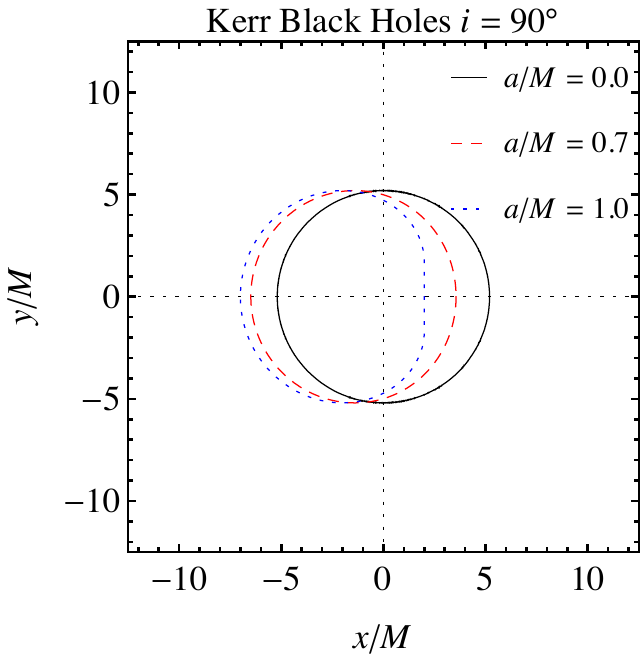}\\
\vspace{0.3cm}
\includegraphics[type=pdf,ext=.pdf,read=.pdf,width=6cm]{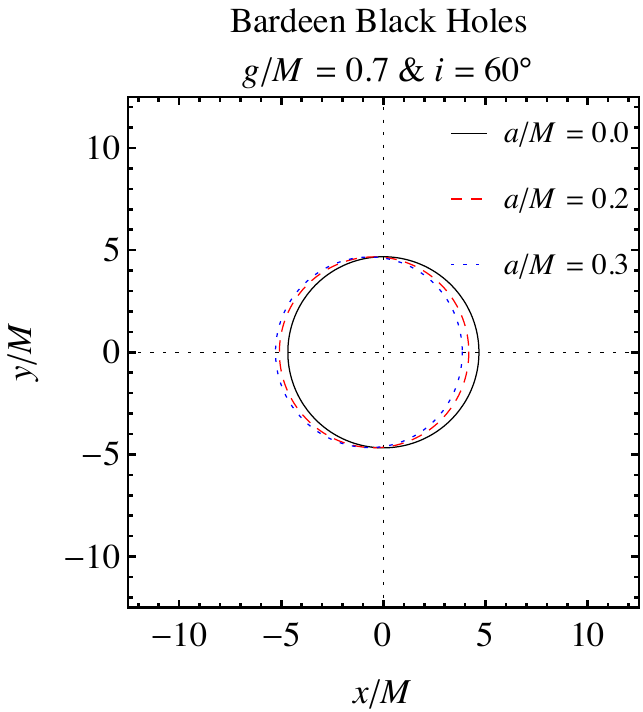}
\hspace{0.5cm}
\includegraphics[type=pdf,ext=.pdf,read=.pdf,width=6cm]{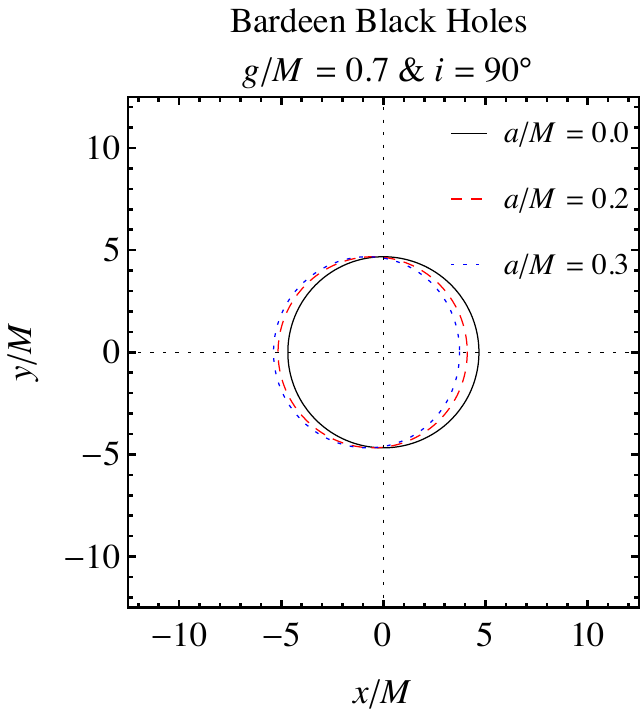}\\
\vspace{0.3cm}
\includegraphics[type=pdf,ext=.pdf,read=.pdf,width=6cm]{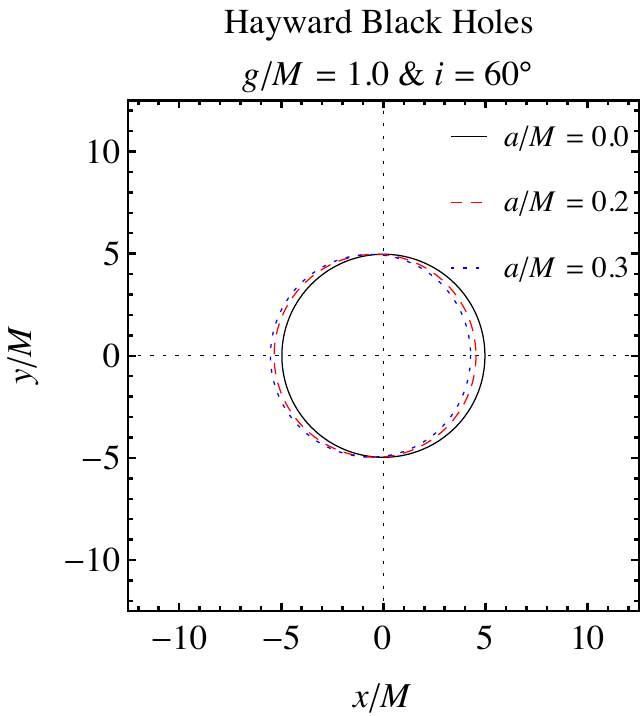}
\hspace{0.5cm}
\includegraphics[type=pdf,ext=.pdf,read=.pdf,width=6cm]{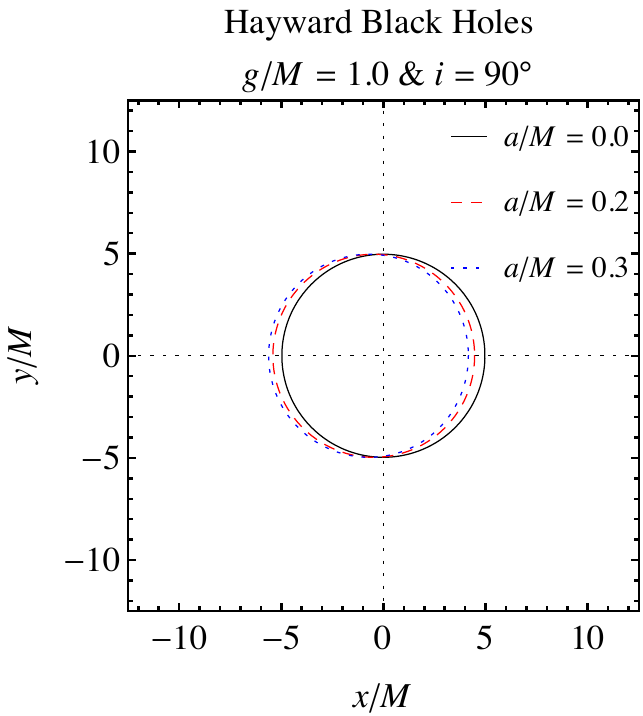}
\end{center}
\vspace{-0.5cm}
\caption{Some examples of boundary of shadow of Kerr BHs (top panels),
Bardeen BHs (central panels), and Hayward BHs (bottom panels) for different
values of the spin parameter $a_*$. The viewing angle is $i = 60^\circ$
(left panels) and $90^\circ$ (right panels).}
\label{fig0}
\end{figure}

\section{Black hole's shadow \label{s-s2}}

The shadow of a BH is a dark area over a bright background appearing in the 
image of an optically thin emitting region around the compact object. The boundary
of the shadow depends only on the geometry of the background and turns out
to correspond to the apparent image of the photon capture sphere as seen by a 
distant observer: if one fires a photon inside the boundary of the shadow, the photon is 
swallowed by the BH; if outside, the photon reaches a minimum distance from 
the compact object and then comes back to infinity. In this section, we will 
briefly review the study of the shadow of a BH (for more details, see e.g. Sec.~63 
of~\cite{book} or Ref.~\cite{Young1976}).

For a photon, the equation of motion for the radial coordinate $r$ in 
Boyer-Lindquist coordinates is
\be
\Sigma^2\bigg(\frac{d r}{d \lambda}\bigg)^2 = \mathcal{R} \, ,   \label{eq-radial}
\ee
where $\lambda$ is an affine parameter, and
\be
\mathcal{R} &=& E^2 r^4+(a^2E^2-L_z^2-\mathcal{Q})r^2 
+2 m [(aE-L_z)^2+\mathcal{Q}]r-a^2\mathcal{Q}\, ,   \label{eq-R} \\
\mathcal{Q} &=& p_\theta^2+\cos^2\theta\bigg(\frac{L_z^2}{\sin^2\theta}-a^2E^2\bigg) \, .
\ee
The parameter $m$ in Eq.~(\ref{eq-R}) is equal to: $M$, for a Kerr BH; $m_{\rm B}$ in
Eq.~(\ref{m-bardeen}) for a Bardeen BH; $m_{\rm H}$ in Eq.~(\ref{m-hayward}), for 
a Hayward BH. $E$, $L_z$, and $\mathcal{Q}$ are constants of motion and are, 
respectively, the energy, the component of the angular momentum parallel to the BH
spin, and the so-called Carter constant. $p_\theta$ is the canonical momentum 
coniugate to $\theta$.

It is convenient to minimize the number of parameters by introducing the variables 
$\xi = L_z/E$ and $\eta = \mathcal{Q}/E^2$. $\xi$ and $\eta$ are very simply related  
to the so-called ``celestial coordinates'' $x$ and $y$ of the image, as seen by an 
observer at infinity who receives the light ray, by
\be
x = \frac{\xi}{\sin i}\, ,  \quad
y = \pm (\eta+a^2\cos^2 i-\xi^2\cot^2 i)^{1/2}\, ,
\ee
where $i$ is the angular coordinate of the observer at infinity. Precisely, $x$
is the apparent perpendicular distance of the image from the axis of symmetry and $y$
is the apparent perpendicular distance of the image from its projection on the equatorial
plane.

The radial equation of motion (\ref{eq-radial}) depends on $\theta$ only in the factor
$\Sigma^2$, and is decoupled from $\phi$ and $t$. Thus the behavior of
$\mathcal{R}(r)$ determines the type of orbit and the question of escape versus plunge
for given $\xi$ and $\eta$. Since motion is only possible when $\mathcal{R}(r) \geq 0$,
the analysis of the position of its roots (especially roots in $r \geq r_+$, where $r_+$ is
the horizon) is a powerful method of investigation of photon orbits. Qualitatively,
there are three kinds of photon orbits:

(i) $\mathcal{R}(r)$ may have no roots in $r \geq r_+$ (capture orbits), in which case the
photon arrives from infinity and then crosses the horizon;

(ii) $\mathcal{R}(r)$ has real roots in $r \geq r_+$ (scattering orbits), in which case the
motion of photon is described by null geodesics which have a turning point $\dot{r}=0$;

(iii) unstable orbits of constant radius, which separate the capture and the scattering
orbits, determined by
\be
\mathcal{R}(r_*)=\frac{\partial \mathcal{R}}{\partial r}(r_*)=0,~~ \mathrm{and}
~~\frac{\partial^2 \mathcal{R}}{\partial r^2}(r_*)\geq0 ,   \label{unstable}
\ee
with $r_*$ being the greatest real root of $\mathcal{R}$.

The apparent shape of the BH can be found by looking for the
unstable orbits. Every orbit can be characterized by the
constants of motion $\xi$ and $\eta$, and the set of unstable
circular orbits ($\xi_c$, $\eta_c$) can be used to plot a closed curve
in the $xy$ plane which represents the boundary of the BH shadow. 
The apparent image of the BH is larger than its geometrical size, 
because the BH bends light rays and thus the actual cross section is
larger than the geometrical one. From Eqs.~(\ref{eq-R}) and (\ref{unstable}),
the equations determining the unstable orbits of constant radius are
\be
\mathcal{R} &=& r^4+(a^2-\xi_c^2-\eta_c)r^2+2m[\eta_c+(\xi_c-a)^2]r  -a^2\eta_c = 0 ,  \nonumber \\
\frac{\partial \mathcal{R}}{\partial r} 
&=& 4r^3+2(a^2-\xi_c^2-\eta_c)r+2m[\eta_c+(\xi_c-a)^2]  
= 0.    \label{eq-critical}
\ee
In the case of a Schwarzschild BH ($a=0$ and $m=M$), the solution is~\cite{book}
\be
\eta_c(\xi_c)=27M^2-\xi_c^2 ,
\ee
so the apparent image of the BH is a circle of radius $\sqrt{27}M$ 
(black solid circles in the left panels of Fig.~\ref{fig0}). For a Kerr BH, one finds
\be
\xi_c~&&=\frac{1}{a(r-M)}[M(r^2-a^2)-r(r^2-2Mr+a^2)] , \nonumber \\
\eta_c~&&=\frac{r^3}{a^2(r-M)^2}[4a^2M-r(r-3M)^2],
\ee
where $r$ is the radius of the unstable orbit. These two equations determine,
 parametrically, the critical locus $(\xi_c,\eta_c)$, which is the set of unstable
 circular orbits. The boundary of the shadows of Kerr BHs with $a/M=0.0$, $0.7$, and 
 $1.0$ are shown in Fig.~\ref{fig0} for an observer with angular coordinate 
 $i = 60^\circ$ (top left panel) and for one on the equatorial plane 
 (top right panel).

In the case of the Bardeen and Hayward BHs, the solutions are more complicated. 
Yet, we can solve Eq.~(\ref{eq-critical}) with $m=m_{\rm B}$ and $m=m_{\rm H}$, 
and obtain the formula of the critical locus $(\xi_c,\eta_c)$ for these metrics. In
Fig.~\ref{fig0}, we show some examples of boundary of shadow of Bardeen BHs 
(central panels, with deformation parameter $g/M=0.7$), and Hayward BHs (bottom 
panels, with deformation parameter $g/M=1.0$) for $a_* = 0.0$, 0.2, and 0.3.
Such a low values of the spin parameter with respect to the Kerr case is motivated 
by the fact that the maximum value of $a_*$ is lower than 1 for $g/M \neq 0$ and 
reduces to 0 for $g/M \approx 0.77$ (Bardeen) and 1.06 (Hayward). For higher 
values of the spin parameter, there is no horizon, and these metrics describe the 
gravitational field around unstable configurations of exotic matter.

\begin{figure*}
\begin{center}
\includegraphics[type=pdf,ext=.pdf,read=.pdf,width=7cm]{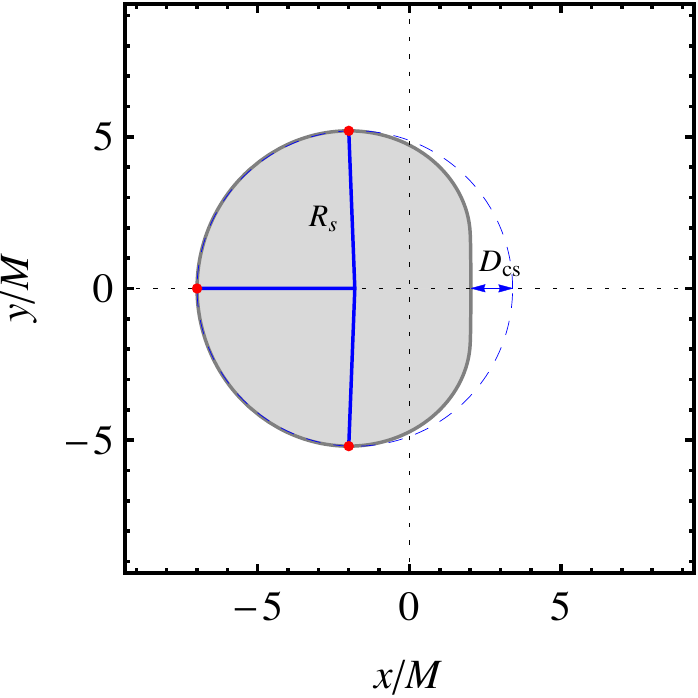}
\hspace{0.5cm}
\includegraphics[type=pdf,ext=.pdf,read=.pdf,width=7cm]{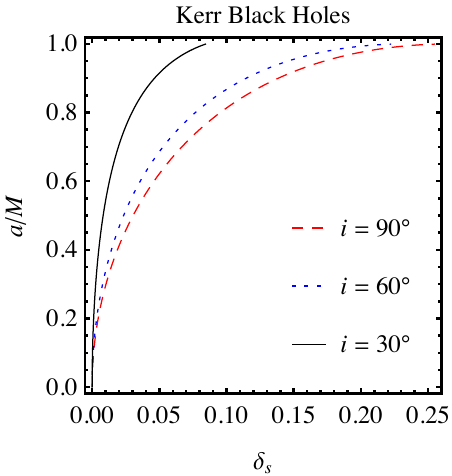}
\end{center}
\vspace{-0.7cm}
\caption{Left panel: BH's shadow with the two parameters that approximately
characterized its shape: the radius $R_s$ (defined as the radius of the circle
passing through the three red points located at the top, bottom, and most
left end of the shadow) and the dent $D_{cs}$ (the difference between the
right endpoints of the circle and of the shadow). Right panel: Spin parameter
$a_* = a/M$ as a function of the distortion parameter $\delta_s = D_{cs}/R_s$
for Kerr BHs and a viewing angle $i = 90^\circ$ (red dashed curve), $60^\circ$
(blue dotted curve), and $30^\circ$ (black solid curve).}
\label{fig1}
\end{figure*}

\begin{figure*}
\begin{center}
\includegraphics[type=pdf,ext=.pdf,read=.pdf,width=15cm]{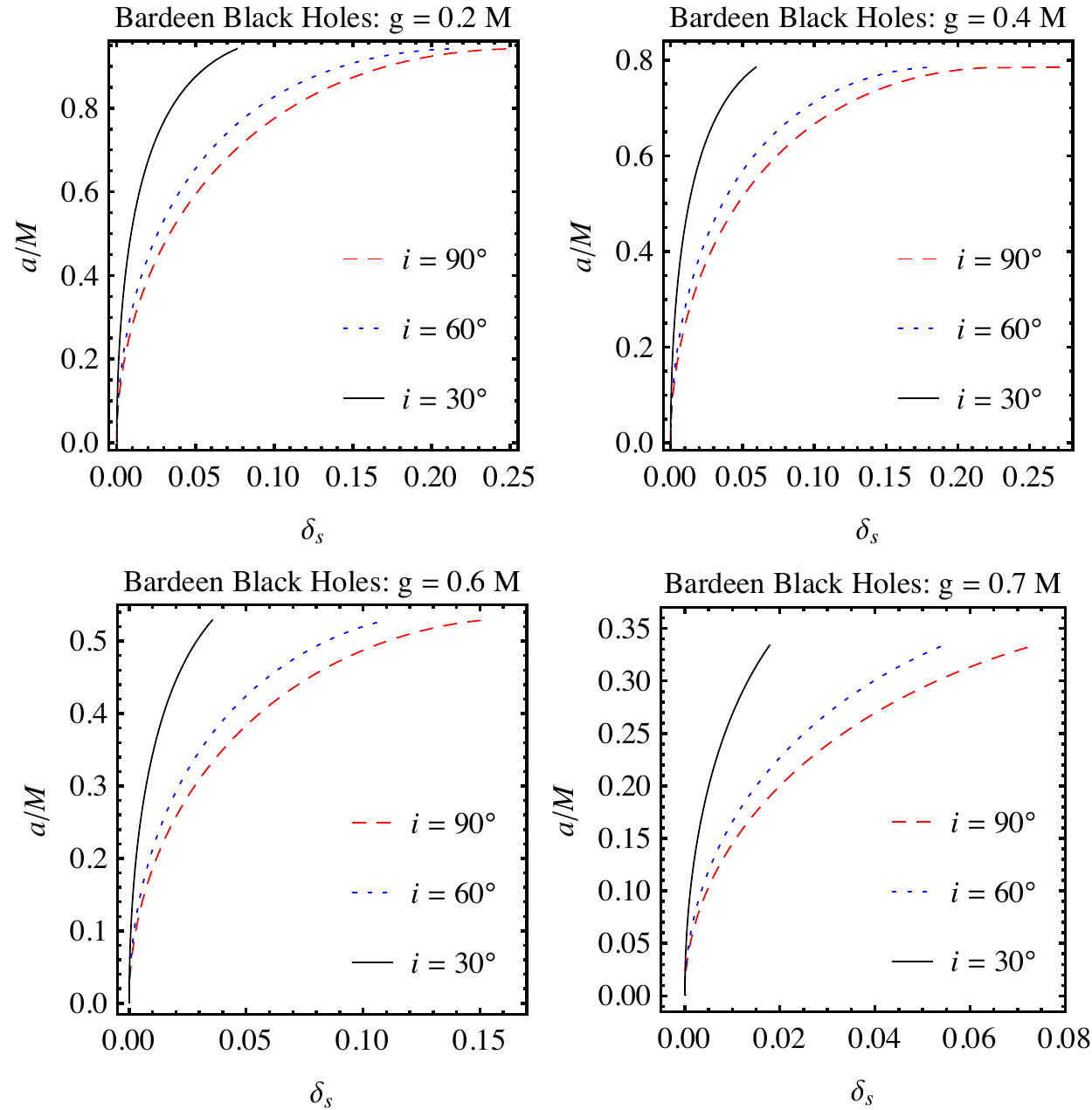}
\end{center}
\vspace{-0.5cm}
\caption{As in the right panel of Fig.~\ref{fig1} for the case of Bardeen BHs
with $g/M = 0.2$ (top left panel), 0.4 (top right panel), 0.6 (bottom left panel),
and 0.7 (bottom right panel).}
\label{fig3}
\end{figure*}

\begin{figure*}
\begin{center}
\includegraphics[type=pdf,ext=.pdf,read=.pdf,width=15cm]{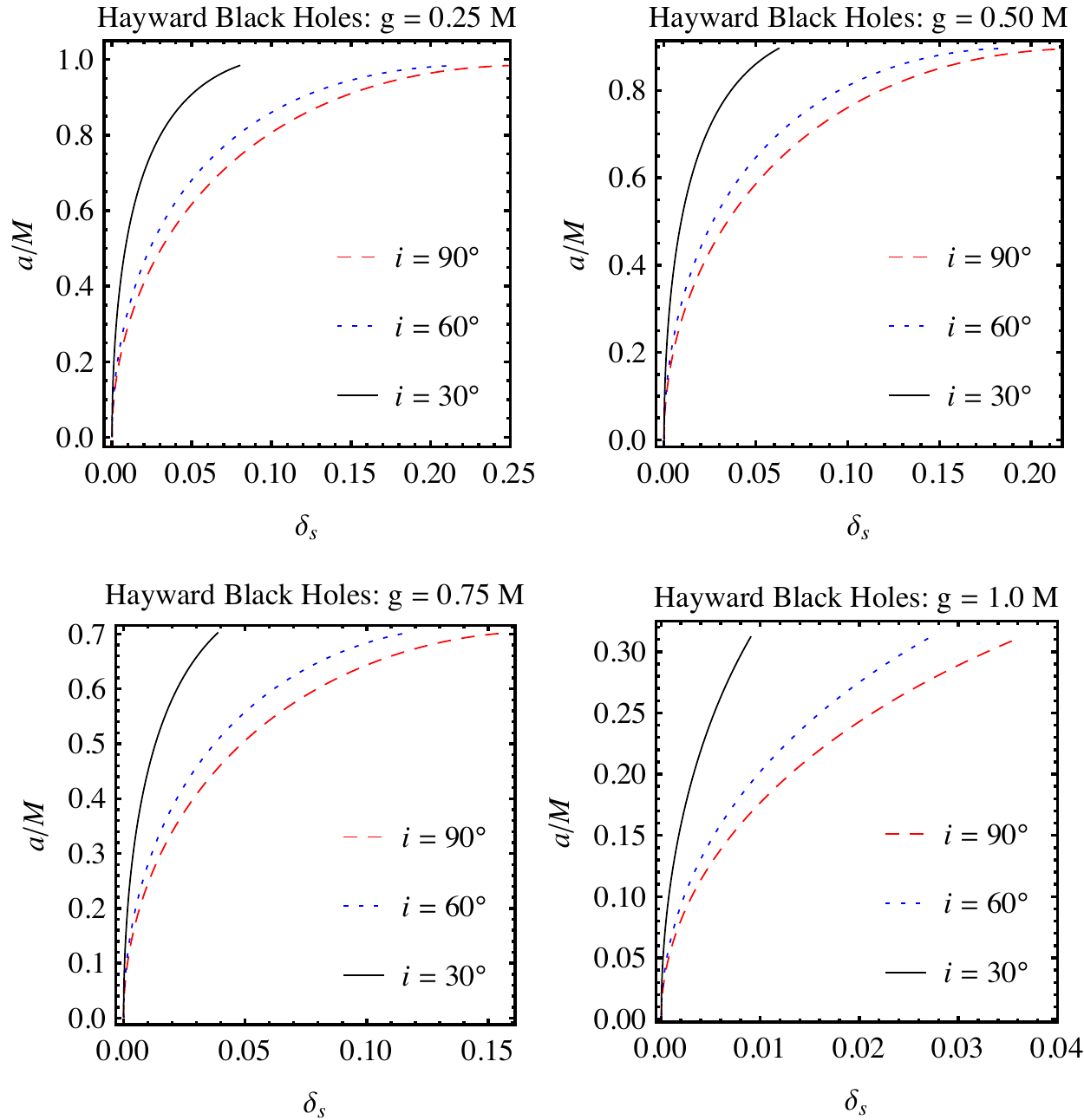}
\end{center}
\vspace{-0.5cm}
\caption{As in the right panel of Fig.~\ref{fig1} for the case of Hayward BHs
with $g/M = 0.25$ (top left panel), 0.50 (top right panel), 0.75 (bottom left panel),
and 1.00 (bottom right panel).}
\label{fig4}
\end{figure*}

\section{Measuring the Kerr spin parameter from the black hole's shadow \label{s-m}}

In the observation of the shadow of a BH, it is helpful to introduce a parameter
that approximately characterizes its shape~\cite{maeda}. At first approximation, the 
shape of the shadow of a BH is a circle, so we approximate the shadow by a circle 
passing through three points, which are located at the top position, the bottom 
position, and the most left end of its boundary (the three red points in the left panel 
of Fig.~\ref{fig1}). The radius $R_{\rm s}$ of the shadow is hereby defined by the 
radius of this circle. On the other hand, when a BH rotates, the difference of the 
photon capture radius between co-rotating and counter-rotating particles introduces 
a dent on one side of the shadow. Unlike in the electromagnetic case, in gravity
the spin-orbit interaction term is repulsive when the orbital angular momentum of the
photon is parallel to the BH spin (the capture radius thus decreases), and attractive
in the opposite case (the capture radius increases). The dent is more pronounced
for fast-rotating objects and it is very clear for the case of an extremal Kerr BH
and a large viewing angle $i$ (the blue-dotted curves in the left panels of Fig.~\ref{fig0}). 
The size of the dent is evaluated by $D_{\rm cs}$, which is the difference between 
the right endpoints of the circle and of the shadow (see the left panel of Fig.~\ref{fig1}). 
Thus the distortion parameter $\delta_{\rm s}$ of the shadow is defined by 
$\delta_{\rm s}=D_{\rm cs}/R_{\rm s}$, which can be adopted as an observable in 
astronomical observations~\cite{maeda}.

In the case of Kerr backgrounds, the exact shape of the shadow depends only on 
the BH spin parameter, $a_*$, and the line of sight of the distant observer with 
respect to the BH spin, $i$. For a given inclination angle $i$, there is a one-to-one
correspondence between $a_*$ and the distortion parameter $\delta_{\rm s}$.
If we have an independent estimate of the viewing angle and we measure the
distortion parameter of the shadow of a Kerr BH, we can infer its spin parameter
$a_*$~\cite{maeda}. The right panel of Fig.~\ref{fig1} shows the curves describing the
spin parameter $a_* = a/M$ as a function of the distortion parameter $\delta_{\rm s}$
for Kerr BHs and an inclination angle $i = 90^\circ$ (red dashed curve), $60^\circ$
(blue dotted curve), and $30^\circ$ (black solid curve).

The same idea can be applied to non-Kerr BHs. If we consider the Bardeen and
Hayward BHs with a specific value of the deformation parameter $g/M$, for a given 
inclination angle
$i$ there is a one-to-one correspondence between the spin $a_*$ and the
distortion parameter $\delta_{\rm s}$. The counterpart of the right panel in Fig.~\ref{fig1}
for the Bardeen and Hayward BHs are shown, respectively, in Fig.~\ref{fig3} and
Fig.~\ref{fig4} for several values of $g/M$. The relation between $a_*$ and $\delta_{\rm s}$
depends on $g/M$ and it reduces to the Kerr one for $g=0$.

\begin{figure*}
\begin{center}
\includegraphics[type=pdf,ext=.pdf,read=.pdf,width=15cm]{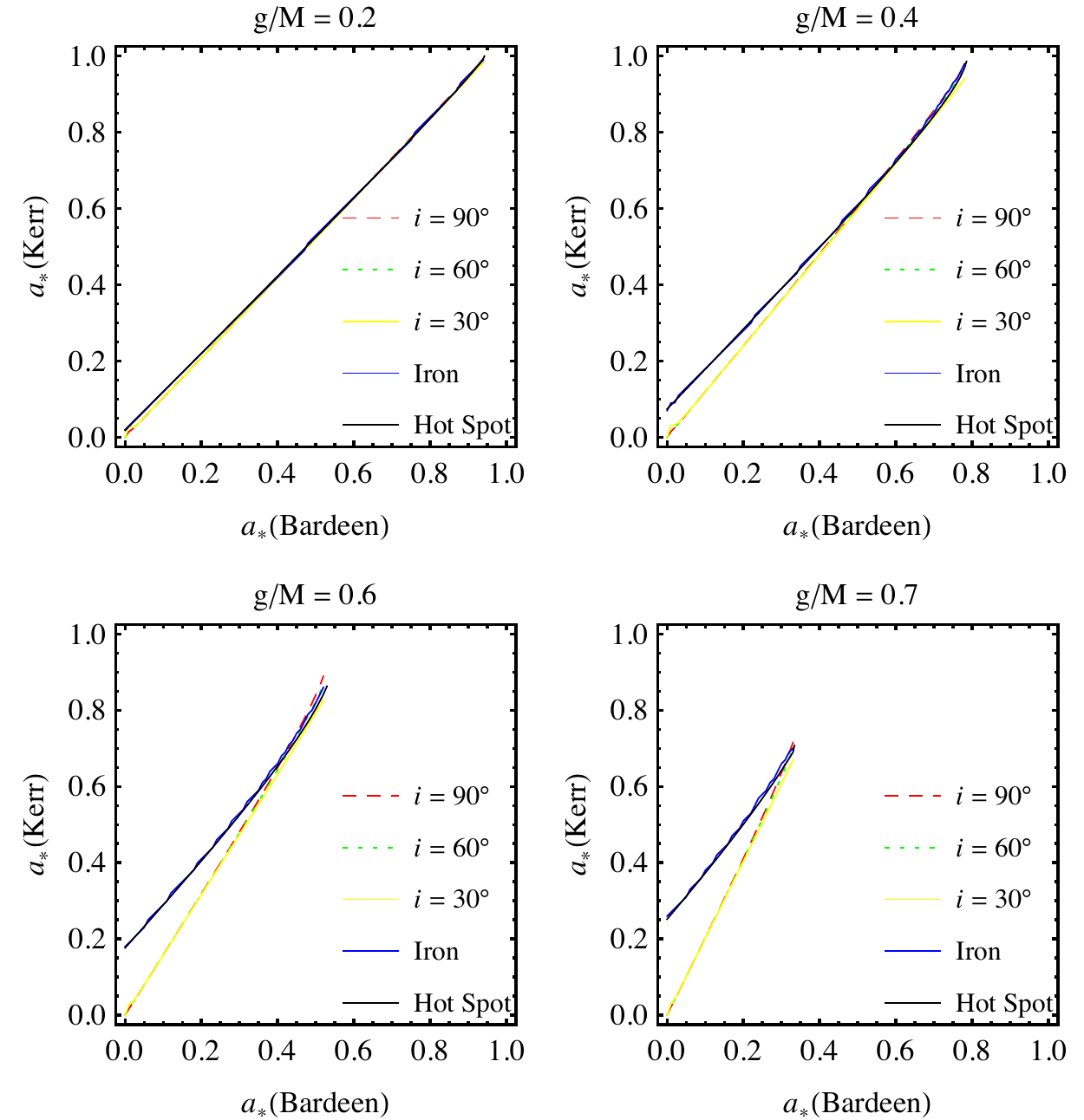}
\end{center}
\vspace{-0.5cm}
\caption{Spin parameter of a Bardeen BH, $a_*({\rm Bardeen})$, against the spin
parameter that one would infer for this object assuming the Kerr background,
$a_* ({\rm Kerr})$, through the determination of the distortion parameter of the
shadow $\delta_s$ (red dashed curve, dotted green curve, and yellow solid
curve respectively for a viewing angle $i=90^\circ$, $60^\circ$, and $30^\circ$),
the analysis of the K$\alpha$ iron line (blue solid curve), and the frequency of
a test-particle at the ISCO radius (hot spot model, black solid curve). $g/M = 0.2$
(top left panel), 0.4 (top right panel), 0.6 (bottom left panel), and 0.7 (bottom
right panel). See the text for more details.}
\label{fig5}
\end{figure*}

\begin{figure*}
\begin{center}
\includegraphics[type=pdf,ext=.pdf,read=.pdf,width=15cm]{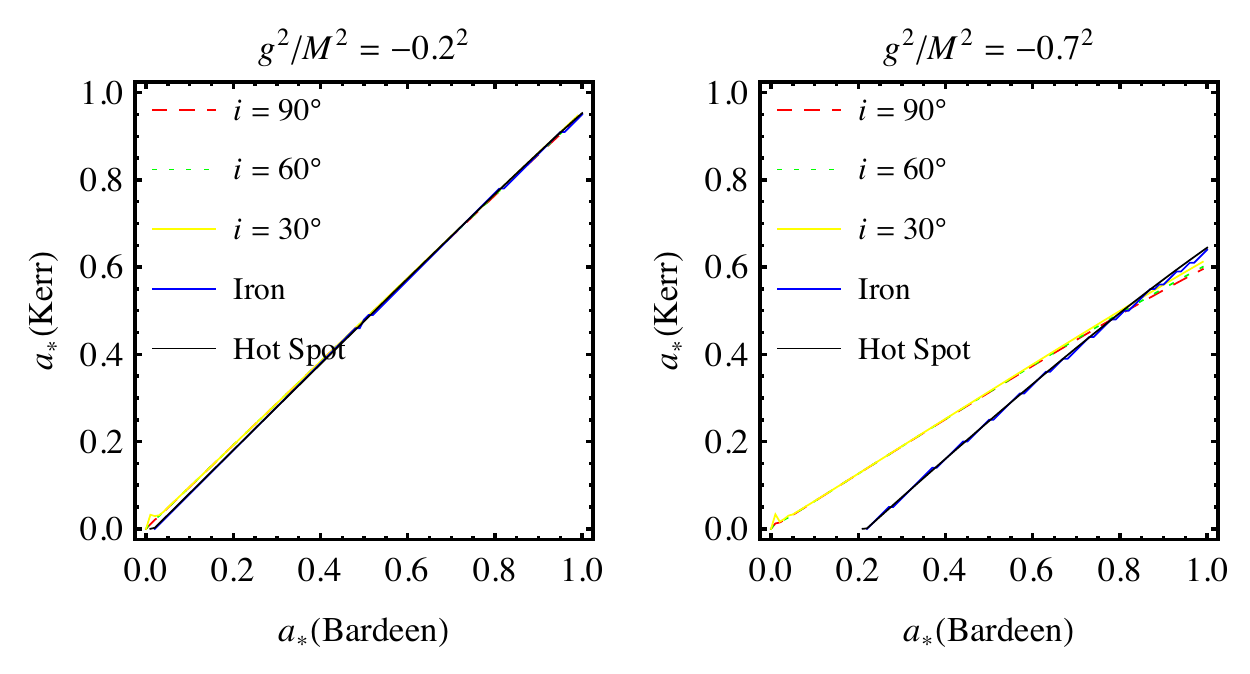}
\end{center}
\vspace{-0.5cm}
\caption{As in Fig.~\ref{fig5} for $(g/M)^2 = - (0.2)^2$ (left panel) and
$-(0.7)^2$ (right panel).}
\label{fig6}
\end{figure*}

As in Ref.~\cite{cb-ci}, we can now address the question of what happens if we
measure the Kerr spin parameter of a non-Kerr BH; that is, we estimate the spin
parameter $a_*$ from the measurement of the distortion parameter $\delta_{\rm s}$
of the shadow of a Bardeen or Hayward BH assuming it is of Kerr type. In the
Kerr metric, there is a one-to-one correspondence between the spin parameter
and $\delta_{\rm s}$ and therefore, from the measurement of the latter, we can
infer the Kerr spin from 
\be
a_*^{\rm Kerr} = a_*^{\rm Kerr}(\delta_{\rm s}) \, . 
\ee
However, in the Bardeen
(and Hayward) background the distortion parameter is given by 
$\delta_{\rm s}(a_*^{\rm B},g/M)$ and therefore the Kerr spin parameter of a Bardeen
BH with spin $a_*^{\rm B}$ and charge $g/M$ is 
\be
a_*^{\rm Kerr} = a_*^{\rm Kerr}[\delta_{\rm s}(a_*^{\rm B},g/M)] \, .
\ee
The result of a similar measurement is reported in Figs.~\ref{fig5} and \ref{fig6} for the case
of Bardeen BHs, and in Fig.~\ref{fig7} for Hayward BHs. The result is independent
of the inclination angle $i$. For non-rotating objects, this approach trivially
provides the correct spin: for any spherically symmetric BH, the shadow is always a
circle, $\delta_{\rm s} = 0$, independently of possible deviations from the Kerr
background. For slow-rotating BHs, we find
a discrepancy between the actual spin parameter of the compact object and the
one inferred assuming a Kerr metric. 
Such a difference increases for higher values of $a_*$ and it can
be quite significant for large values of $g/M$.

\section{Discussion \label{s-m2}}

The distortion parameter $\delta_{\rm s}$ is just a number and therefore
cannot determine both the exact spin $a_*$ and the deformation $g/M$. 
However, it is remarkable that the shadow of a regular BH with a large $g/M$ 
cannot mimic the one of a fast-rotating Kerr BH. In other words, if we observe
a shadow that looks like the one of a Kerr BH with high spin, we can constrain
the deformation parameter $g/M$. In general, this is not possible, and we
should combine this measurement with an independent one, in order to break
the degeneracy between $a_*$ and $g/M$.

In Ref.~\cite{cb-ci}, one of us has
shown that the simultaneous measurement of the Kerr spin via the continuum-fitting
and the iron line methods cannot fix this problem for Bardeen BHs. 
The two techniques provide the same information on the geometry of the
spacetime around the compact object. We
can now check if the combination of the Kerr spin parameter measured with the shadow 
approach can be combined with another estimate and if it is possible to
distinguish a true Kerr BH from a Bardeen or Hayward one. Near future VLBI 
facilities will be able to image only the shadow of super-massive BH candidates.
The analysis of the K$\alpha$ iron line is currently the only technique that
can provide a relatively robust estimate of the Kerr spin parameter of these
objects (the continuum-fitting method can be used only for stellar-mass BH 
candidates). It is supposed that VLBI experiments will be able to observe
also hot blobs of plasma orbiting around nearby super-massive BH candidates 
and infer their angular frequency at the ISCO radius. Such a time measurement
can provide an independent measurement of the Kerr spin parameter, as in
the Kerr metric there is a one-to-one correspondence between BH spin and
angular frequency of the ISCO.

The profile of the K$\alpha$ iron line depends on the background metric, 
the geometry of the emitting region, the disk emissivity, and the disk's 
inclination angle with respect to the line of sight of the distant observer.
In the Kerr spacetime, the only relevant parameter of the background 
geometry is the spin parameter $a_*$, while $M$ sets the length of the 
system, without affecting the shape of the line. In those sources for which 
there is indication that the line is mainly emitted close to the compact 
object, the emission region may be thought to range from the radius of 
the ISCO, $r_{\rm in} = r_{\rm ISCO}$, to some outer radius $r_{\rm out}$. 
In principle, 
the disk emissivity could be theoretically calculated. In practice, that is 
not feasible at present. The simplest choice is an intensity profile 
$I_{\rm e} \propto r^{\alpha}$ with index $\alpha < 0$ to be determined 
during the fitting procedure. The fourth parameter is the inclination of the 
disk with respect to the line of sight of the distant observer, $i$. The 
dependence of the line profile on $a_*$, $i$, $\alpha$, and $r_{\rm out}$ 
in the Kerr background has been analyzed in detail by many authors, 
starting with Ref.~\cite{iron-p1}. In the case of non-Kerr backgrounds,
see e.g. Ref.~\cite{cb-iron-p1}.

\begin{figure*}
\begin{center}
\includegraphics[type=pdf,ext=.pdf,read=.pdf,width=15cm]{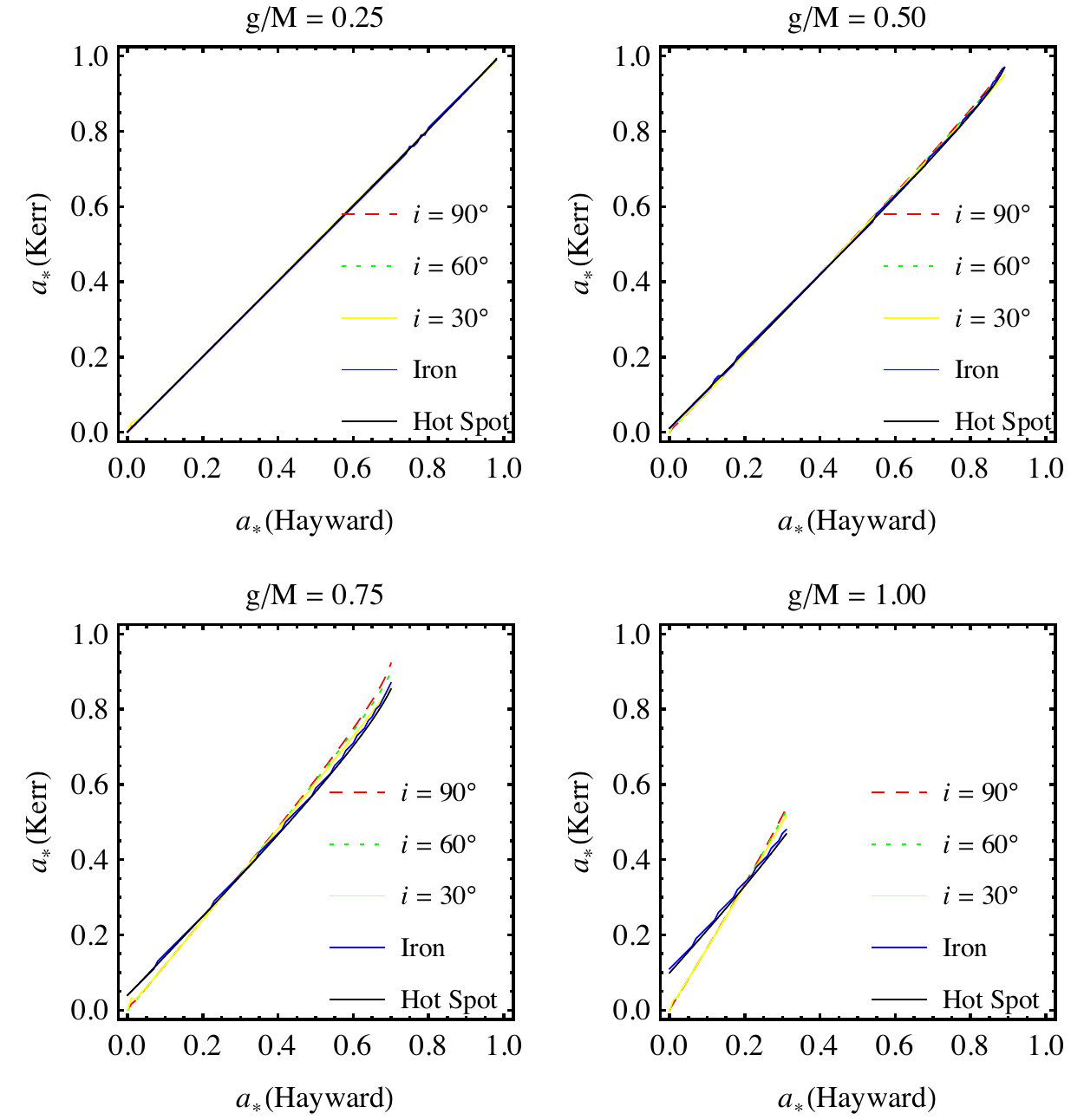}
\end{center}
\vspace{-0.5cm}
\caption{As in Fig.~\ref{fig5} for Hayward BHs with $g/M = 0.25$ (top left panel),
0.50 (top right panel), 0.75 (bottom left panel), and 1.00 (bottom right panel).}
\label{fig7}
\end{figure*}

The profile of the K$\alpha$ iron line can be obtained by computing 
the photon flux number density measured by a distant observer, that is
\be
N_{E_{\rm obs}} &=& \frac{1}{E_{\rm obs}} 
\int I_{\rm obs}(E_{\rm obs}) d \Omega_{\rm obs} =
\frac{1}{E_{\rm obs}} \int w^3 I_{\rm e}(E_{\rm e}) 
d \Omega_{\rm obs} \, ,
\ee
where $I_{\rm obs}$ and $E_{\rm obs}$ are, respectively, the specific intensity 
of the radiation and the photon energy as measured by the distant observer, 
while $I_{\rm e}$ and $E_{\rm e}$ are the same quantities in the rest-frame of 
the emitter. $d\Omega_{\rm obs}$ is the solid angle seen by the distant observer
and $w$ is the redshift factor
\be\label{eq-red}
w = \frac{E_{\rm obs}}{E_{\rm e}} =  
\frac{k_\alpha u^{\alpha}_{\rm obs}}{k_\beta u^{\beta}_{\rm e}}\, .
\ee
Here $k^\alpha$ is the 4-momentum of the photon, $u^{\alpha}_{\rm obs} = (-1,0,0,0)$ 
is the 4-velocity of the distant observer, and 
$u^{\alpha}_{\rm e} = (u^t_{\rm e},0,0, \Omega u^t_{\rm e})$ is the 4-velocity of 
the emitter. $\Omega$ is the angular velocity for equatorial circular orbits.
$I_{\rm e}(E_{\rm e})/E_{\rm e}^3 = I_{\rm obs} (E_{\rm obs})/E_{\rm obs}^3$ 
follows from the Liouville's theorem. Using the normalization condition
$g_{\mu\nu}u^{\mu}_{\rm e}u^{\nu}_{\rm e} = -1$, one finds
\be
u^t_{\rm e} = - \frac{1}{\sqrt{-g_{tt} - 2 g_{t\phi} \Omega - g_{\phi\phi} \Omega^2}} \, .
\ee
The redshift factor is thus given by
\be\label{eq-red-g}
w = \frac{\sqrt{-g_{tt} - 2 g_{t\phi} \Omega - g_{\phi\phi} \Omega^2}}{1 + 
\lambda \Omega} \, ,
\ee
where $\lambda = k_\phi/k_t$ is a constant of the motion along the photon path.
Doppler boosting, gravitational redshift, and frame dragging are entirely encoded 
in the redshift factor $w$.
As the K$\alpha$ iron line is intrinsically narrow in frequency, we 
can assume that the disk emission is monochromatic (the rest frame energy 
is $E_{\rm{K}\alpha} = 6.4$~keV) and isotropic with a power-law radial 
profile:
\be
I_{\rm e}(E_{\rm e}) \propto \delta (E_{\rm e} - E_{\rm{K}\alpha}) r^{\alpha} \, .
\ee

Let us now consider the possibility that an astrophysical BH candidate is a
Bardeen BH and that we want to measure the spin parameter of this object
with the K$\alpha$ iron line analysis, assuming that the object is a Kerr BH. 
In this case, we can use an approach similar to the one discussed in 
Ref.~\cite{cb-iron-p1} and define the reduced $\chi^2$:
\be
\chi^2_{\rm red} (a_*^{\rm Kerr}, i)
&=& \frac{1}{n} \sum_{i = 1}^{n} \frac{\left[N_{i}^{\rm Kerr} 
(a_*^{\rm Kerr}, i) - N_{i}^{\rm B}
(a_*^{\rm B}, g/M, i^{\rm B}) 
\right]^2}{\sigma^2_i} \, ,
\label{eq-chi2-ka}
\ee 
where the summation is performed over $n$ sampling energies $E_i$ and
$N_i^{\rm Kerr}$ and $N_i^{\rm B}$ are the normalized photon 
fluxes in the energy bin $[E_i,E_i+\Delta E]$,
respectively for the Kerr and the Bardeen metric. The error $\sigma_i$ is 
assumed to be 15\% the photon flux $N_i^{\rm B}$, which is roughly the 
accuracy of current observations in the best situations. In this paper, all the 
calculations are done with an intensity profile index $\alpha = -3$ and an 
outer radius $r_{\rm out} = r_{\rm in} + 100 \, M$. For specific values of 
$a_*^{\rm B}$ and $g/M$, we can find the minimum of the reduced $\chi^2$, 
and we thus obtain what we call the Kerr spin parameter.

In the case of the frequency of a hot spot at the ISCO radius, the idea is
the same. In the Kerr background, there is a one-to-one correspondence
between $\Omega_{\rm ISCO}$ and $a_*$, so we can write 
$\Omega_{\rm ISCO}^{\rm Kerr} (a_*^{\rm Kerr})$ and the inverse function 
$a_*^{\rm Kerr}(\Omega_{\rm ISCO}^{\rm Kerr})$. In the Bardeen (and Hayward)
background the frequency at the ISCO radius depends on both the spin $a_*^{\rm B}$ 
and the deformation parameter $g/M$, so 
$\Omega_{\rm ISCO}^{\rm B} (a_*^{\rm B},g/M)$. If we measure the
frequency of a hot spot at the ISCO radius and we assume that the object is a Kerr
BH, while it is of Bardeen type, one finds:
\be
a_*^{\rm Kerr} = a_*^{\rm Kerr}[\Omega_{\rm ISCO}^{\rm B} (a_*^{\rm B},g/M)] \, ,
\ee
which is the Kerr spin parameter of the Bardeen BH.

Figs.~\ref{fig5}-\ref{fig7} show also the possible
measurements of the Kerr spin parameter with these two techniques (blue solid 
line for the iron line, black solid line for the hot spot). First, it seems 
like the iron line and hot spot approaches provide essentially the same result. This
is because the two techniques are essentially sensitive to the properties at the 
ISCO radius. Second, when we compare the results of these approaches with 
the measurements inferred from the shadow, we see that the measurements
disagree in the case of non-rotating BHs, while the discrepancy decreases
as the spin parameter increases and there is almost no difference for near extremal 
states. So, the determination of the distortion parameter of the shadow of a
BH candidate can potentially test the nature of the compact object when combined 
with the iron line analysis or the hot spot approach for non-rotating or slow-rotating 
objects. The degeneracy between $a_*$ and $g/M$ cannot be solved in the case
of near extremal BHs, which confirms the difficulties, found in Ref.~\cite{cb-ci}, to 
test this kind of non-Kerr metrics. Lastly, let us notice that here we have always 
considering an ``ideal'' observation, neglecting possible uncertainty in the 
measurements. We thus adopt an optimistic point of view, and the difficulties to 
measure deviations from the Kerr solution are even more challenging.

\section{Summary and conclusions \label{s-c}}

Astrophysical BH candidates are thought to be the Kerr BHs of general
relativity, but the actual nature of these objects is still to be verified. The
analysis of the thermal spectrum of thin accretion disks and of the profile
of the K$\alpha$ iron line are today the only available approaches to probe
the spacetime geometry around BH candidates and test the Kerr BH
hypothesis. However, there is a strong correlation between the spin parameter
and possible deviations from the Kerr solution and it is not possible to
check the nature of a specific source with a single measurement. As shown
in Ref.~\cite{cb-ci}, at least for some non-Kerr metrics, the
disk's thermal spectrum and the iron line provide essentially the same
information. So, the two measurements may be consistent with the ones of
a Kerr BH with a certain value of the spin parameter $a_*$ even if the
object is actually something else with a different spin. The
combination of the two measurements does not break the degeneracy.

In this paper, we have investigated the possibility of measuring the Kerr
spin parameter from the BH shadow. Near future mm/sub-mm VLBI
facilities will be able to observe the emitting region around nearby
super-massive BH candidates with a resolution comparable to their
gravitational radius. If the gas is geometrically thick and optically thin,
we will observe a dark area over a brighter background. While the
intensity map of this image depends on the kind of accretion process
and the emission mechanisms, the boundary of the shadow is
completely determined by the geometry of the spacetime. The shape of
the shadow is expected to be a circle for non-rotating BHs or a viewing
angle $i = 0^\circ$ or $180^\circ$, and slightly deformed otherwise, as
a consequence of the coupling between the spin of the compact object
and the photon angular momentum. Such a deformation can be measured
in terms of the distortion parameter $\delta_s$, which, in turn, may
provide an estimate of the spin parameter $a_*$.

If the compact object is not a Kerr BH, but we assume it is, this technique
still provides the correct value of $a_*$ for non-rotating objects, but a
wrong measurement for near extremal states. We have compared these
measurements with the ones of the spin that one could infer
from the iron line and from the determination of the frequency at the
ISCO radius (a kind of measurement that should become possible and
reliable with VLBI facilities). For non-rotating BHs, the shadow approach
would provide a different result, so the possible inconsistency between two
measurements may be an indication of deviations from the Kerr solution.
All the approaches converge instead to the same value for near extremal
BHs. This work thus confirm the intrinsic difficulty to test the Kerr-nature
of astrophysical BH candidates, even with future facilities. It may however be
possible that good measurements of the shadow, in which one can extract
more than one parameter characterizing the shape, are able to break
the degeneracy between the spin and possible deviations from the Kerr
geometry for any value of $a_*$.


\begin{acknowledgments}
This work was supported by the NSFC grant No.~11305038, the Shanghai 
Municipal Education Commission grant for Innovative Programs No.~14ZZ001,
the Thousand Young Talents Program, and Fudan University.
\end{acknowledgments}


\end{document}